\documentclass[preprint,showpacs,showkeys]{elsarticle}

\usepackage{gensymb}
\usepackage[T1]{fontenc}
\usepackage{bm}
\usepackage{graphicx}
\usepackage{amssymb}
\usepackage{amsfonts}
\usepackage{amsmath}
\usepackage{multirow}
\usepackage{pstricks}
\usepackage{graphicx}

\begin{document}

\begin{frontmatter}

\title{Influence of ZnTe based distributed Bragg reflectors on the yellow range luminescence of self assembled CdTe QDs}

\author[add1]{J.-G.~Rousset\corref{cor}}

\author[add1]{J.~Kobak}

\author[add1]{E.~Janik}

\author[add2]{M.~Parlinska-Wojtan}

\author[add1]{T.~Slupinski}

\author[add1]{J.~Suf\mbox{}fczy\'{n}ski}

\author[add1]{A.~Golnik}

\author[add1]{P.~Kossacki}

\author[add1]{M.~Nawrocki}

\author[add1]{W.~Pacuski}

\address[add1]{Institute of Experimental Physics, Faculty of Physics, University of Warsaw, ul. Pasteura 5, PL-02-093 Warszawa, Poland}
\address[add2]{Institute of Nanotechnology, University of Rzeszow, 35-959 Rzeszow, Poland}

\cortext[cor]{Corresponding author:\\
email: j-g.rousset@fuw.edu.pl\\
phone: +48 22 55 32 707}



\begin{abstract}
The influence of a distributed Bragg reflector composed of ZnTe, MgTe, and MgSe superlattices on photoluminescence of self assembled CdTe quantum dots (QD) emitting in the yellow spectral range is investigated. In the case of QDs grown on a distributed Bragg reflector the photoluminescence intensity is enhanced by more than one order of magnitude, whereas the single QD lines are broadened as compared to the case of QDs grown on a ZnTe buffer. Structural and chemical analysis reveal an unintentional formation of a thin ZnSe layer induced by the growth interruption needed for the deposition of the QDs sheet. Sharp emission lines from individual quantum dots are recovered in the case of a thicker ZnTe layer grown prior to the QDs. This indicates that growth interruptions might be responsible for the QD emission line broadening.
\end{abstract}



\end{frontmatter}
\maketitle


\section{Introduction}

State of the art distributed Bragg reflectors (DBR) exhibit an optical reflectivity much higher than metallic mirrors, what makes them a perfect component for implementation in advanced photonic devices and semiconductor spectroscopy studies. For example, planar microcavities\cite{Kavokin_MC2007} involving DBRs are well proven to enhance a semiconductor optical features thanks to the increased light-matter interaction strength. Recently it has been shown that a DBR can be used to enhance the magneto-optical Kerr effect in a dilute magnetic semiconductor layer.\cite{Koba_JEWA2013, Koba_EPL2014} A DBR designed for a maximum reflectivity in a range of wavelengths centred around $\lambda_0$ consists of a stack of $\lambda_0 / 4n_i$ thick layers with different refractive indices $n_i$. Microcavities defined by such DBRs based on III-V compounds have been extensively developed and studied in the past twenty years.\cite{Weisbuch_PRL92, Deng_Science2002, Lagoudakis_PRL2003, Houdre_PSSB2005, Amo_nature2009, Tanese_PRL2012} On the other hand, II-VI semiconductors exhibit higher exciton binding energies than III-V materials. Therefore, II-VI based microcavities are expected to enable functional semiconductor devices and the fundamental investigations of the light-matter coupling at higher temperatures than their III-V counterparts.\cite{Dang98, guo00, Tawara00, Kruse04, Kruse_APL_2008, Pac_APL2009, Jakubczyk_APPA2009, Sebald_APL12, Rou_JCG2013}

Furthermore, II-VI compounds based emitters are good candidates to solve the problem of the low efficiency of III-V based emitters in the green-yellow spectral range.\cite{Kram_JDispTech2007,Rousset_JCG2014} In particular, accomplishment of yellow range (wavelength between $560$ and $595\ nm$) optoelectronic devices could have important applications in medicine, data transmission through optical fibers, or in optical spectroscopy. In this context, self assembled CdTe quantum dots (QDs) in a ZnTe barrier designed to emit in the yellow spectral range are particularly desired. Besides, CdTe QDs doped with a single magnetic ion may find applications in the promising field of solotronics (an optoelectronics based on solitary dopants)\cite{Koen_NatureMat2011,Kobak_NComms2014} or in quantum information storage and processing\cite{michler_book2003,LeGall_PRL_2009,Goryca_PRL2009} schemes. In the case of such applications and fundamental investigations, the figure of merit is the emission pattern of a single quantum dot. To facilitate these investigations, one needs to enhance the emission of a given QD covering a relatively wide spectral range ($\approx 10\ nm$). On the opposite, for yellow optoelectronics applications or for the control of the spatial, energy and temporal dynamics properties of QDs emission\cite{Dousse_Nature2010,Jomek_APL2012,Jomek_JAP2013,Pac_CGandD2014,Jomek_ACSnano2014} a microcavity with a spectrally sharp mode is needed.

In this work, we present results on the growth of self assembled CdTe QDs emitting in the yellow range. The deposition of the QDs sheet on a DBR is shown to strongly enhance the QD emission intensity in a wide spectral range (beneficial for fundamental investigations on QDs, solotronics and QD based information processing). The growth of QDs embedded in a microcavity structure exhibiting a sharp mode required for optoelectronics applications is also demonstrated.

\section{Structures: growth and preparation}

\begin{figure}
    \includegraphics[width=1\linewidth]{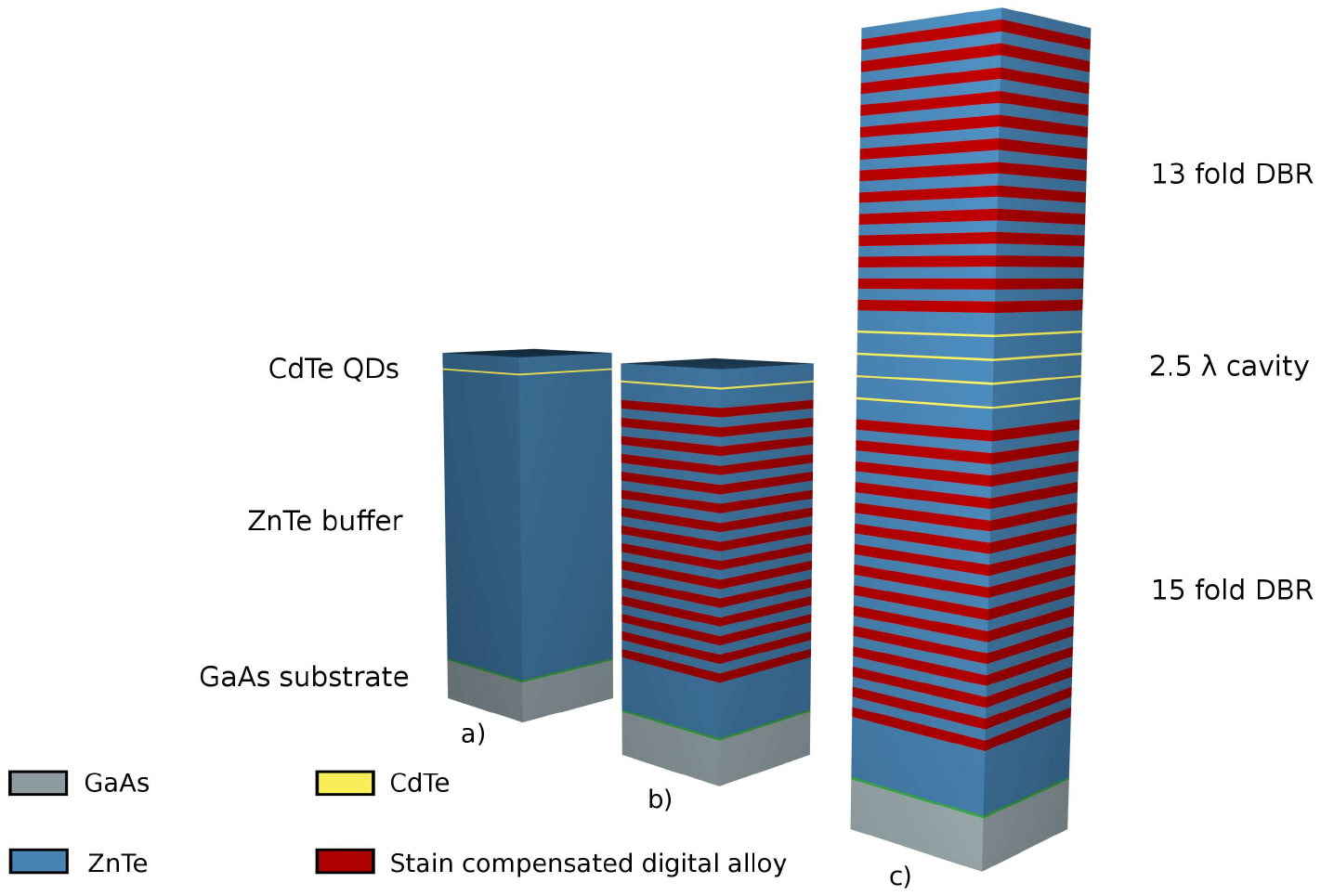}
  \caption{Structures. a) CdTe QDs on ZnTe buffer without any photonic structure, b) half-cavity structure: a single QDs sheet is embedded in a ZnTe barrier deposited on a $10$ fold DBR. The thickness of the ZnTe barrier is set to form a $\lambda$ cavity which top reflector is the ZnTe / air interface. c) full cavity structure: the $4$ QDs sheets are embedded in a $2.5\ \lambda$ cavity (1 QDs sheet at each antinode of the electromagnetic standing wave).}
  \label{struc}
\end{figure}

\begin{figure}
    \includegraphics[width=1\linewidth]{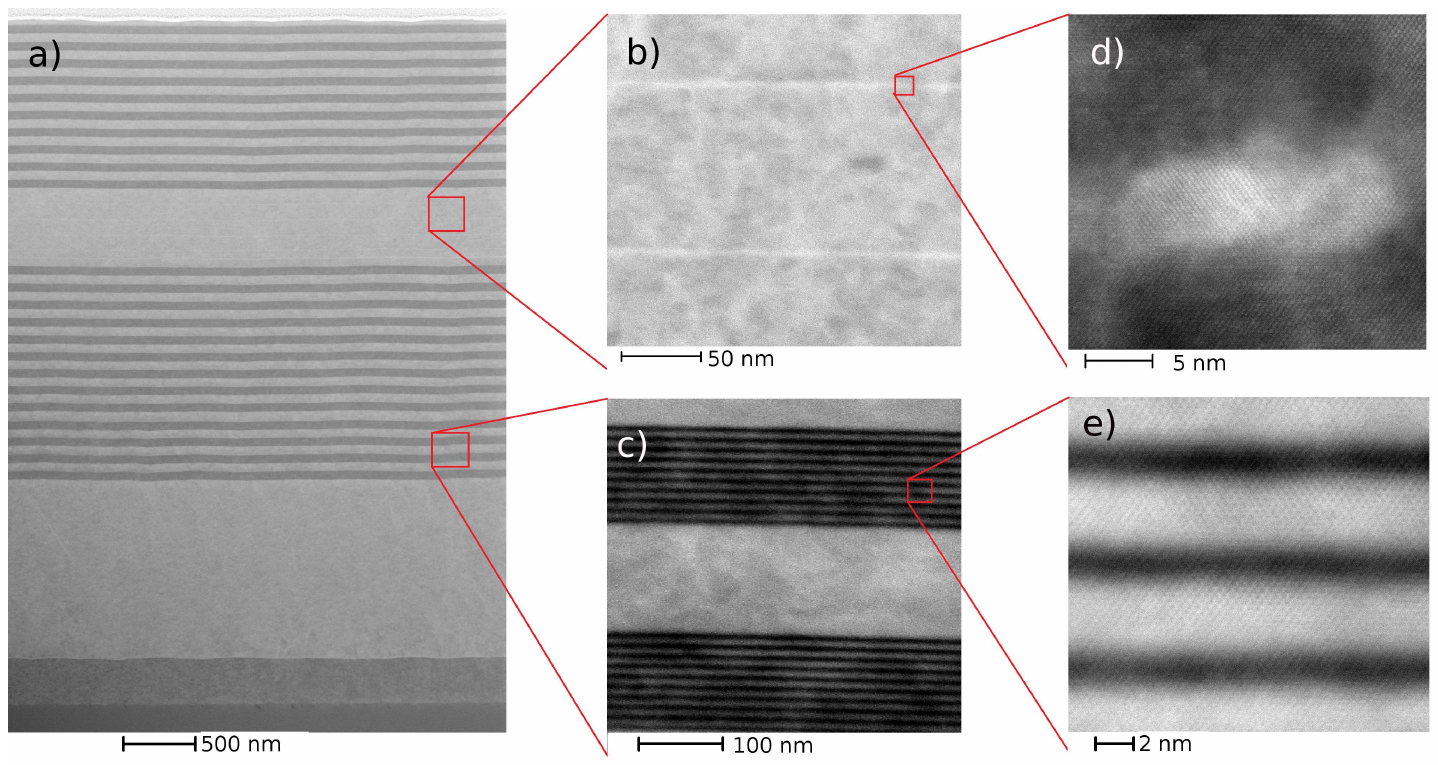}
  \caption{High annular angular dark field (HAADF) imaging of a microcavity structure in the scanning transmission electron microscopy (STEM) configuration: a) whole microcavity structure with the top and bottom DBR, b) close up on the QDs sheets, c) close up on the DBR layers. The layers constituting the short period superlattice (low refractive index digital alloy) are clearly visible. d) close up on a single CdTe QD: on this high resolution image, the crystaline structure of the QD is observed. e) close up showing the steep interfaces between the layers of the short period superlattice and the good crytal quality.}
  \label{HAADF}
\end{figure}

The structures are grown by molecular beam epitaxy (MBE) on ($100$) oriented GaAs substrates. Three different structures are grown and studied: (i) a reference structure consisting of CdTe/ZnTe QDs on a $2\ \mu m$ ZnTe buffer without any photonic structure (Fig. \ref{struc} a)). (ii) The half-cavity structure consisting of a $\approx \ 1\mu m$ ZnTe buffer, overgrown by a $10$ fold DBR, and with a single QDs sheet deposited at the center of a $\lambda$ cavity. The ZnTe / air interface plays the role of top reflector for the cavity (see Fig. \ref{struc} b)). (iii) The full-cavity structure consisting of a $\approx \ 1\mu m$ ZnTe buffer on which a cavity between two DBRs is grown: respectively $13$ fold for the bottom DBR and $10$ fold for the top one. Four QDs sheets are deposited in the $2.5 \lambda$ cavity (Fig. \ref{struc} c)) using the same growth parameters as for the structure (ii). The microcavities studied comprise DBRs lattice matched to ZnTe. The high refractive index layers within the DBR are ZnTe layers. The low refractive index layers are obtained by growing a short period, strain compensated ZnTe|MgSe|ZnTe|MgTe superlattice (see Fig. \ref{HAADF} c) and e)) resulting in a digital alloy.\cite{Pac_APL2009} The substrate temperature for the growth of the DBRs is $T_{DBR}=367\degree C$, whereas the self organized QDs are obtained by the deposition of $3$ monolayers of CdTe at a substrate temperature of $T_{QD}=356\degree C$. In order to limit the growth interruptions and shorten the overall growth process of cavities embedding numerous QDs sheets, the QDs formation was not induced by the well established method involving amorphous tellurium\cite{Tinjod_APL2003} deposition. The CdTe layer was overgrown without any growth interruptions by ZnTe barrier resulting in the formation of QDs with a relatively high planar density\cite{Karczewski_APL1999,Kruse_Nanotech2011,Kobak_APPA2011,Kobak_JCG2013}. The high annular angular dark field (HAADF) imaging in the scanning transmission electron microscopy (STEM) allows to observe the crystalline structure of the QDs and barrier (see Fig. \ref{HAADF} d)).

\section{Photoluminescence}

\begin{figure}
    \includegraphics[width=0.8\linewidth]{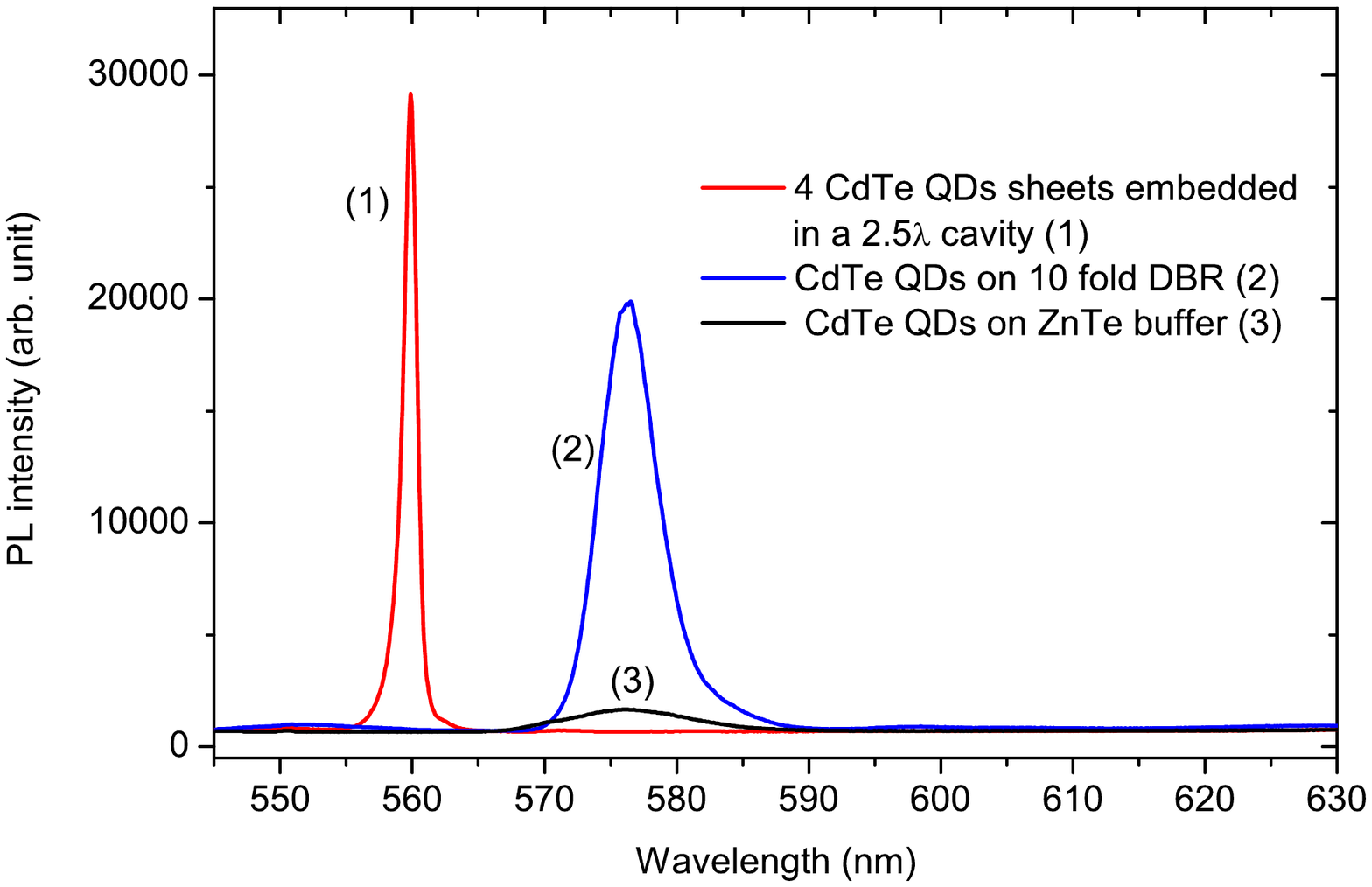}
  \caption{Macro-photoluminescence of the QDs embedded to the full cavity structure (1), the QDs grown on a DBR (2), and the reference QDs grown on a ZnTe buffer (3). The PL intensity from the QDs ensemble is enhanced by more than an order of magnitude by the half-cavity structure involving the DBR.}
  \label{macroPL}
\end{figure}

\begin{figure}
    \includegraphics[width=0.8\linewidth]{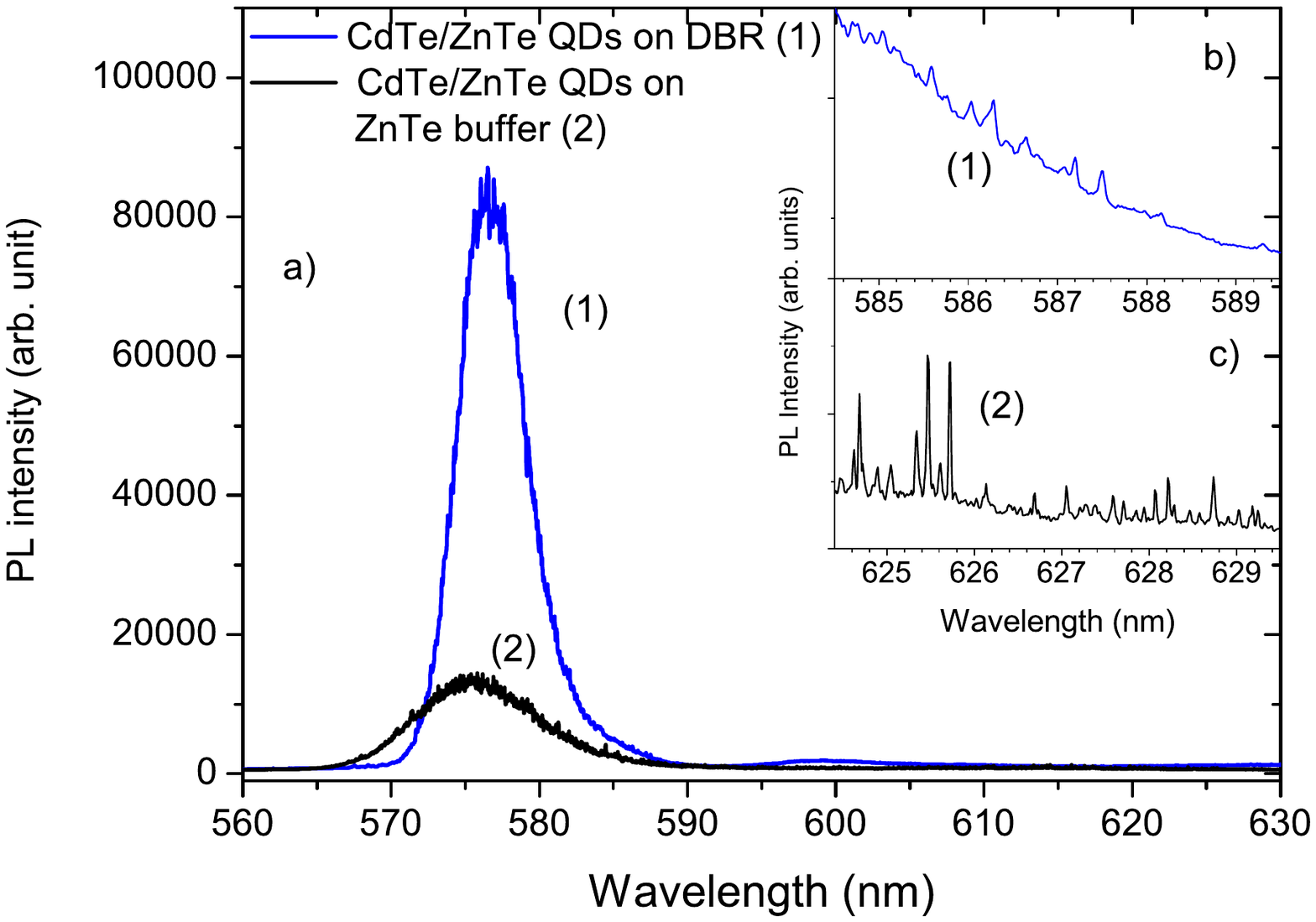}
  \caption{Micro-photoluminescence of the QDs grown on the DBR (1) and of the reference QDs on ZnTe buffer (2). The CdTe QDs grown on the DBR exhibit broadened single emission lines as compared to the QDs on ZnTe buffer [insets b) and c)].}
  \label{microPL}
\end{figure}

The samples are characterized by macro-photoluminescence (macro-PL) measurements (diameter of the laser spot on the sample surface: $\approx 50\ \mu m$) to compare the studied structures emission intensities (Fig. \ref{macroPL}) and by micro-photoluminescence ($\mu$-PL) measurements (size of the laser spot on the sample: $\approx 1\ \mu m$) to observe single QDs emission lines (Fig. \ref{microPL}). The QDs are excited above the barrier by a continuous wave laser beam at $\lambda_{exc}=532\ nm$.

As seen in Fig. \ref{macroPL}, the PL intensity of the QDs ensemble in the yellow range ($\lambda_{peak} \approx 575\ nm$) grown on a ten fold DBR is $13$ times higher than in the case of QDs grown on a $2\ \mu m$ ZnTe buffer. This intensity enhancement is accounted for a higher extraction coefficient of the light emitted by the QDs resulting from the coupling of the emission to the increased density of electromagnetic states (mode) of the cavity defined by the bottom DBR and the ZnTe / air interface. In the case of QDs on a ZnTe buffer, most of the light emitted towards the substrate or back reflected from the cap / air interface is absorbed by the substrate.

Also in the case of the full cavity structure the macro-PL measurements show a significant enhancement of the PL intensity from the QDs (Fig.\ref{macroPL}). The interpretation here is, however, less straightforward. The cavity quality factor is higher as compared to the case of the half-cavity (it is high enough to result in mode linewidth narrower than the QD ensemble emission peak, in contrary to the case of the  half-cavity). On the one hand, this increases the effect of cavity enhanced extraction and may account for a net Purcell effect on the QD emission. On the other hand, the excitation is less efficient due to absorption of the excitation beam in the top DBR. At the same time, not a single QD sheet, but  $4$ QDs sheets (excited with an unequal efficiency) contribute to the collected signal. The overall emission intensity is a result of the interplay of the above mentioned factors.

\begin{figure}
    \includegraphics[width=0.8\linewidth]{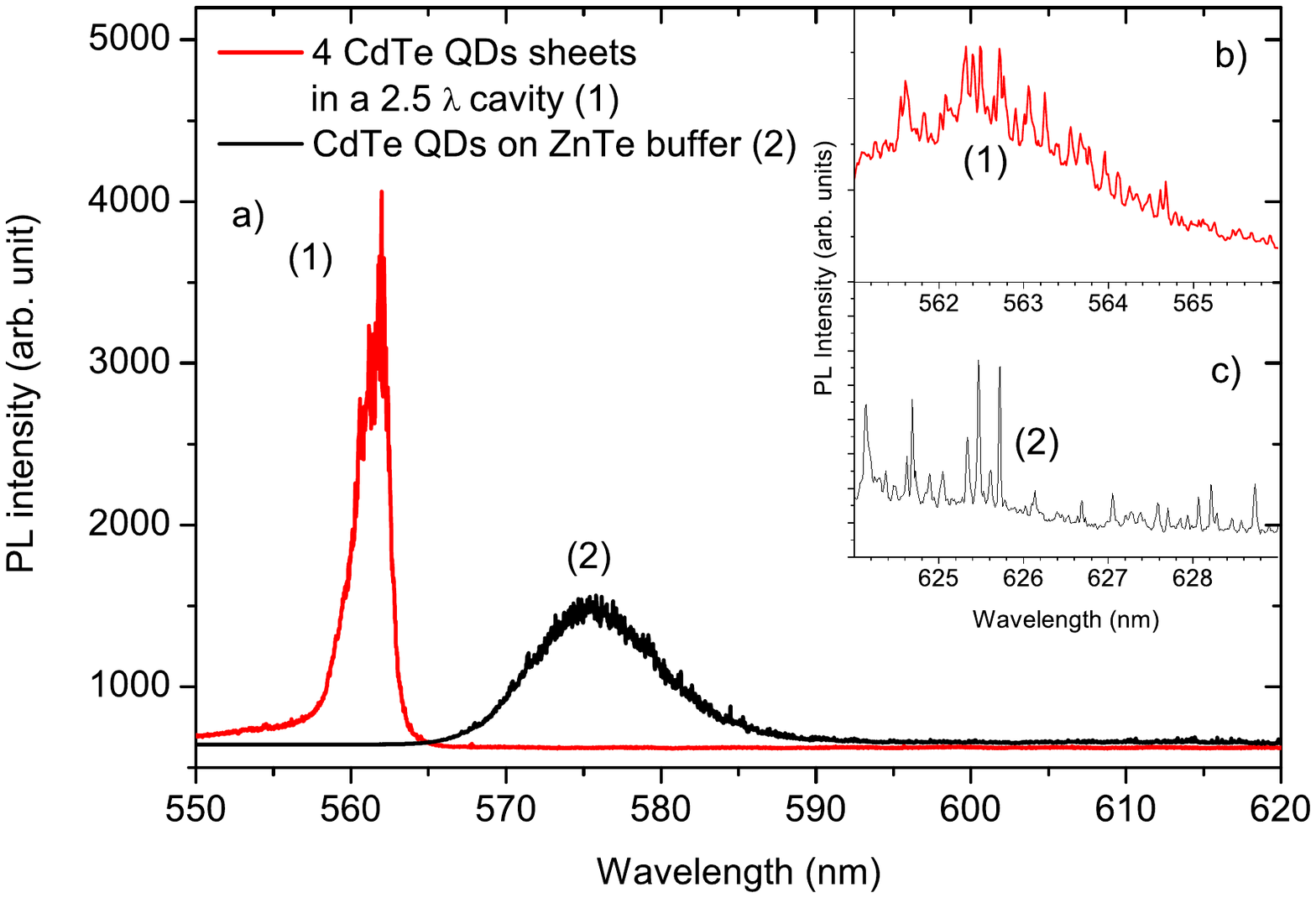}
  \caption{Micro-photoluminescence of the full cavity structure (1) and of the reference QDs on a ZnTe (2). As seen in the close up [insets b) and c)], the single QDs lines exhibit a FWHM comparable in the case of both structures.}
  \label{MC_DBR}
\end{figure}

The micro-PL measurements reveal that single emission lines of the QDs grown on the DBR are strongly broadened as compared to the QDs grown on the ZnTe buffer (Fig. \ref{microPL}). As seen in the inset of Fig.\ref{MC_DBR}, sharp PL lines related to individual QDs are observed for the full-cavity structure although the same procedure for the deposition the CdTe QDs was used for both structures. The cause of the observed broadening can be of structural type (\emph{e. g.}, the propagation of stacking faults through the QDs sheet), of crystalline type (\emph{e. g.}, poor crystalline quality of the QDs or matrix in their vicinity) or of electronic type (\emph{e. g.}, accidental doping). In the next paragraph, we show that the latter effect is predominant.

\section{Structural and chemical analysis}

\begin{figure}
    \includegraphics[width=1\linewidth]{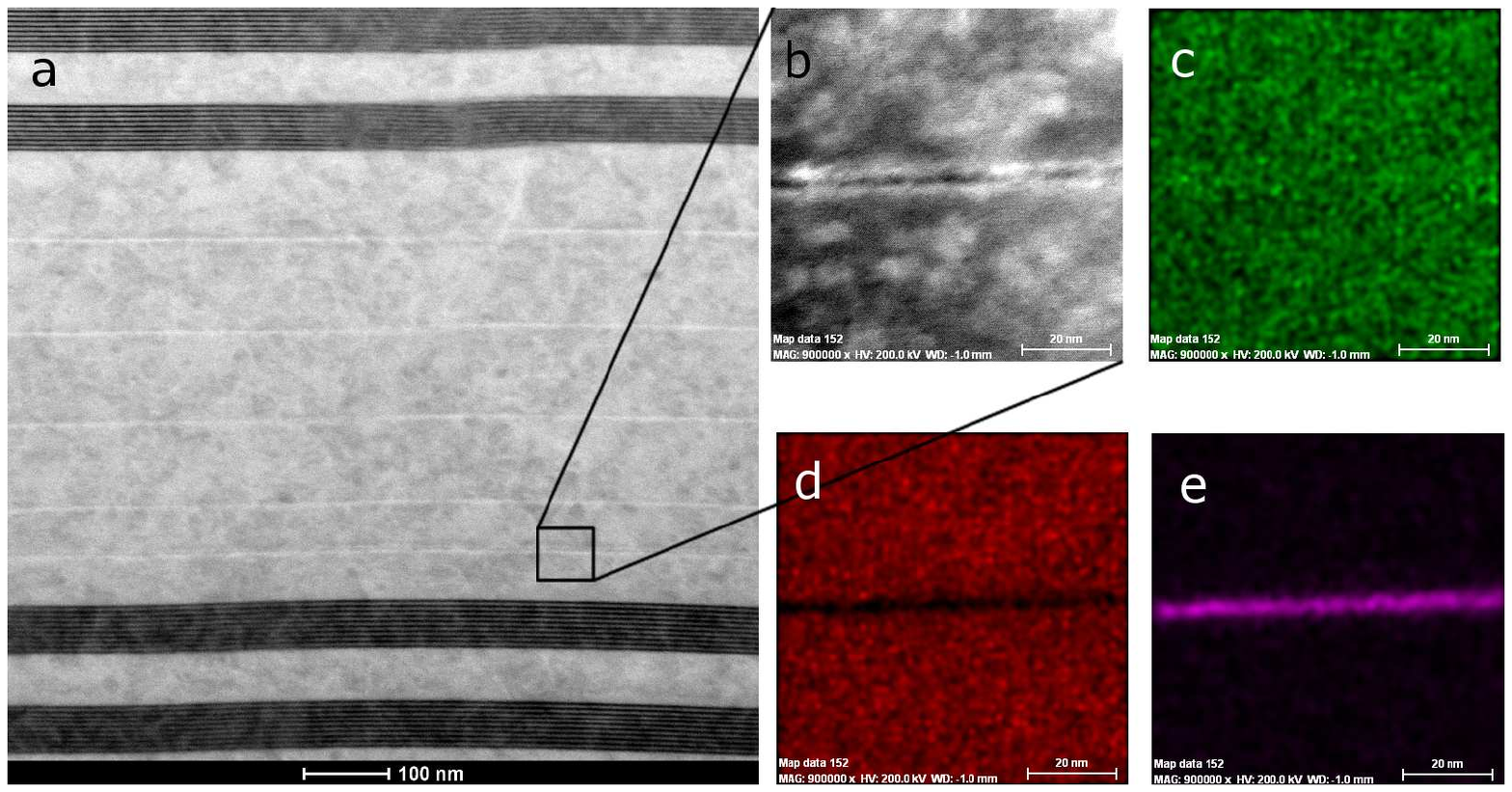}
  \caption{STEM HAADF imagining of the cavity- a. The four QDs sheets are clearly visible, there is no visible stacking fault across the sample and no amorphous inclusion.  However, we distinguish an additional layer under the QDs sheets  - close-up - b. The EDX chemical analysis of the close-up region enables to map the presence of Zn - c, Te - d, and Se - e. This unequivocally shows that a thin ZnSe ($\approx 1\ nm$ thick) layer has been unintentionally deposited during the growth interruption necessary for the growth of the CdTe QDs.}
  \label{HRTEM}
\end{figure}

To identify the origin of the QDs single lines broadening observed for the structure with a single QDs sheet embedded in a half-cavity, imaging with the high angular annular dark field detector (HAADF) in the  scanning transmission (STEM) mode and energy dispersive X-ray spectroscopy (EDX) analysis have been undertaken on a microcavity sample embedding four QDs sheets. As shown in Fig. \ref{HRTEM} a) and b) the the HAADF images do not show any structural defects across the cavity or DBRs, like stacking faults or amorphous inclusions. This indicates a good crystalline quality of the sample. However, an additional layer under the QDs sheets can be seen (Fig. \ref{HRTEM} a, and close-up b). In this region, the presence of Zn, Te and Se is mapped by EDX chemical analysis (Fig. \ref{HRTEM} respectively c, d, e). The mapping shows unequivocally that a thin ZnSe layer ($\approx 1\ nm$ thick) has formed. The unintentional deposition of this ZnSe layer occurs while cooling the substrate from $T_{DBR}=367\deg$ (growth of the DDBR) to $T_{QD}=356\deg$ (growth of the QDs). Because of the higher sticking coefficient of Se atoms with respect to Te atoms\cite{Takafumi_JCG1978}, the residual Se pressure in the growth chamber (even with the Se shutter closed) is sufficient to result in preferable incorporation of Se in place of Te when cooling the substrate.

\section{Discussion}

The recovery of sharp single QDs lines in the case of the full cavity structure indicates that the effect of the intercalated Se in the form of a ZnSe thin layer in the cavity is a short range type and that it affects only the first QDs sheet grown.
Since no structural defects are observed in the vicinity of the unintentional ZnSe layer, we infer that the main influence of this additional layer is a short range, local modification of the electronic environment affecting the optical properties of the nearest QDs sheet. The non negligible thickness of the ZnSe layer ($\approx1\ nm$), being still under the critical relaxation thickness ($\approx 2\ nm$ for ZnSe on ZnTe\cite{Dosho_JAP89, Pearsall_90}) suggests that also a local strain could be involved in the broadening of the single QDs lines observed for the half-cavity structures with a single QDs sheet. The ZnSe layer unintentionally forms during the growth interruption introduced to lower the temperature of the substrate prior to the growth of QDs. Thus, in order to assure a good QDs optical performance it is crucial to avoid interruptions during the growth. This is especially important in the case of the CdTe QDs embedded in the microcavity. This indicates that in growth processes combining selenium and tellurium compounds it is better to grow the CdTe QDs without inducing their formation by amorphous tellurium. The fact that for the half-cavity structure broadened single QDs emission lines are observed together with an enhancement of the PL from the QDs ensemble suggests that the unintentional ZnSe layer does not introduce any non-radiative exciton recombination channel.

\section{Conclusion}

In this work, we show the enhancement of the yellow emission intensity of self assembled CdTe / ZnTe QDs embedded in microcavities defined by DBRs involving selenium and tellurium compounds. The extraction coefficient of the light emitted by the QDs ensemble is $13$ times higher for half-cavity structures as compared to QDs grown on a buffer. In such cavities the spectrally wide cavity mode is well suited for fundamental investigations on QDs, and for applications like solotronics or information storage and processing for which one focuses on the full emission pattern of individual QDs. We show that the formation of a ZnSe layer (identified by EDX analysis) in the ZnTe cavity during the cooling of the substrate for the growth of the CdTe QDs, does not affect the structural and crystalline properties of the sample, as shown by HAADF STEM imaging. However, this additional ZnSe layer may induce a short range modification of the electronic and strain environment affecting the optical properties of the QDs in its proximity (e.g., broadening of individual QD emission lines). Since the influence of these effects is of short range type, the growth of a thicker ZnTe layer after cooling of the substrate and before the deposition of the QDs allows to cancel them. For the full cavity structure, we show the enhancement of the QDs ensemble emission without any effects of single QDs lines broadening.

\section*{Acknowledgments}

This work was partially supported by the National Center for Research and Development in Poland (project LIDER), the Polish National Science Center under decisions DEC-2011/02/A/ST3/00131, DEC-2011/01/N/ST3/04536, DEC-2012/05/N/ST3/03209, DEC-2013/09/B/ST3/02603, DEC-2013/10/E/ST3/00215, UMO-2012/05/B /ST7/02155,  and by the Foundation for Polish Science. Research was carried out with the use of CePT, CeZaMat and NLTK infrastructures financed by the European Union - the European Regional Development Fund within the Operational Programme "Innovative economy" for 2007-2013.


\end{document}